\documentclass{emulateapj}
\usepackage{psfig}
\slugcomment{Draft: \today}
\shorttitle{8~o'clock~Arc}
\shortauthors{Allam et al.}
\begin{document}
\title{The 8~o'clock Arc: 
A Serendipitous Discovery of a Strongly Lensed Lyman Break Galaxy 
in the SDSS DR4 Imaging Data}
\author{
Sahar S.\ Allam\altaffilmark{1,2},
Douglas L.\ Tucker\altaffilmark{1},
Huan Lin\altaffilmark{1},
H.\ Thomas Diehl\altaffilmark{1},
James Annis\altaffilmark{1},
Elizabeth J.\ Buckley-Geer\altaffilmark{1},
Joshua A.\ Frieman\altaffilmark{1,3}}
\altaffiltext{1}{Fermi National Accelerator Laboratory,   P.O. Box 500, Batavia, IL 60510}
\altaffiltext{2}{University of Wyoming, Dept.\ of Physics \& Astronomy, P.O.Box 3905, Laramie, WY 82071}
\altaffiltext{3}{Department of Astronomy and Astrophysics, University of Chicago, 5640 South Ellis Avenue, Chicago, IL 60637}
\begin{abstract}

We report on the serendipitous discovery  of the brightest Lyman Break
Galaxy (LBG)  currently known,  a galaxy   at $z=2.73$  that  is being
strongly  lensed by  the  $z=0.38$  Luminous   Red Galaxy (LRG)   SDSS
J002240.91+143110.4.  The arc of this gravitational lens system, which
we have dubbed the ``8 o'clock arc'' due to its time of discovery, was
initially   identified in the  imaging data  of  the Sloan Digital Sky
Survey  Data   Release  4 (SDSS  DR4);  followup   observations on the
Astrophysical Research Consortium (ARC) 3.5m telescope at Apache Point
Observatory confirmed the lensing nature of this system and led to the
identification of the arc's spectrum as that of an LBG.  The arc has a
spectrum and a redshift remarkably   similar to those of the  previous
record-holder for brightest LBG  (MS 1512-cB58, a.k.a ``cB58''),  but,
with an    estimated     total  magnitude  of    ($g$,$r$,$i$)     $=$
(20.0,19.2,19.0)    and    surface            brightness            of
($\mu_{g}$,$\mu_{r}$,$\mu_{i}$)    $=$    (23.3,  22.5,      22.3) mag
arcsec$^{-2}$,  the 8 o'clock arc is  thrice as bright.  The 8 o'clock
arc, which consists of  three lensed images  of the LBG, is 162\arcdeg
(9.6\arcsec) long  and has a  length-to-width ratio  of 6:1.  A fourth
image of the LBG --- a counter-image --- can also be identified in the
ARC~3.5m $g$-band images.  A simple lens model for the system assuming
a singular isothermal ellipsoid potential yields an Einstein radius of
$\theta_{\rm  Ein}=2.91\arcsec\pm0.14\arcsec$, a   total mass for  the
lensing  LRG (within  the  10.6$\pm0.5$ $h^{-1}$  kpc enclosed  by the
lensed images)     of  $1.04\times10^{12} h^{-1}  M_{\sun}$,    and  a
magnification factor  for   the LBG of  $12.3^{+15}_{-3.6}$.   The LBG
itself is intrinsically quite luminous  ($\approx 6 \times L_{*}$) and
shows indications of massive recent star formation, perhaps as high as
160 $h^{-1} M_{\sun}$ yr$^{-1}$.

\end{abstract}
\keywords{gravitational lensing --- galaxies: high-redshift}
\section{Introduction} 

Strongly lensed galaxies are particularly useful for studies of galaxy
evolution  due to  the magnification  of the  galaxy  magnitude: since
surface  brightness  is conserved by   lensing, the stretching  of the
galaxy  shape increases  the   apparent  brightness   of the    source
galaxy.  These apparently brighter  objects  are then prime candidates
for detailed follow-up studies  at a  fraction  of the telescope  time
that would be necessary for comparable but unlensed galaxies.

If a  strongly lensed galaxy also happens  to be  a Lyman Break Galaxy
(LBG),  so much the  better.  LBGs are  galaxies in which the low-flux
region  of the spectrum blueward   of the Ly$\alpha$ Hydrogen line  at
1216\AA\ has  been redshifted into   the  $U$ band;  LBG samples  thus
provide a vital  window  into  the  galaxy  populations of   the  high
redshift ($z>2.7$) Universe (e.g.,   Adelberger  et al.\ 1998,   2003;
Steidel et al.\  1998; Giavalisco et  al.\ 1998).  LBGs,  however, are
generally  rather faint, and   detailed studies of these high-redshift
galaxies  profit from  the   additional magnification provided for  by
strong lensing (Nesvadba et al.\ 2006).

Previously,  just two examples  of  strongly  lensed  LBGs have   been
discovered: MS1512-cB58 at $z=2.7$ (a.k.a, ``cB58''; Yee et al.\ 1996,
Teplitz et al.\ 2000,  Pettini et al.\ 2002, Baker  et al.\  2004) and
the 1E0657-56 arc+core at $z=3.2$ (Mehlert et al.\ 2001).  A search by
Bentz et al.\ (2003) using  the Sloan Digital   Sky Survey Early  Data
Release (SDSS EDR; Stoughton et al.\ 2002) yielded six bright ($r \sim
20$) candidate LBGs with $z=2.45-2.80$, but  these were later found to
be unlensed bright quasars (Ivison et al.\ 2005).

Here, we report on the serendipitous discovery in the SDSS data of the
brightest case of  these rare objects, a  strongly lensed $z=2.73$ LBG
which we have nicknamed the ``8 o'clock arc''.

This letter  is organized as follows: \S~\ref{sec:DISCOVERY} describes
the initial  discovery   in the SDSS  imaging  data, \S~\ref{sec:conf}
describes  the  confirmatory followup imaging   and spectroscopy,  and
\S~\ref{sec:model} describes modeling  and  comparison with previously
known high redshift LBGs.  \S~\ref{sec:discussion} contains discussion
and \S~\ref{sec:conclusions} concludes. Throughout,  we assume  a flat
cosmology  with $\Omega_{\rm  m}=  0.3$, $\Omega_{\lambda}=0.7$,   and
$H_0=100h$~km~s$^{-1}$~Mpc$^{-1}$, unless otherwise noted.

\section{The Initial Discovery}\label{sec:DISCOVERY}

The  SDSS (York et al.\  2000) is a  digital imaging and spectroscopic
survey  that,  over the course  of five  years, has  mapped nearly one
quarter of  the   celestial  sphere  in  five  filter  bands ($ugriz$;
Fukugita et al.\ 1996 ) down to $r=22.2$ and  has obtained spectra for
$\approx10^6$ astronomical  objects (Adelman-McCarthy et  al.\ 2007) .
Although  the SDSS  completed its  first phase  of  operations in June
2005, a three-year extension known  as SDSS-II is  in progress.   (For
more details on SDSS-II, please consult {\tt www.sdss.org}.)
 
To explore the  effects of interactions  on the properties of galaxies
in different environments, Allam et al.\ (2004) extracted a catalog of
interacting/merging galaxy pairs from  the SDSS imaging  data.  During
visual inspection of  a new version this  catalog (Allam et al.\ 2007)
based upon the SDSS DR4 (Adelman-McCarthy  et al.\ 2006) imaging data,
a very unusual merging galaxy  pair with a galaxy-galaxy separation of
4.02\arcsec was discovered (see Fig.\  1).  The two components of this
system are SDSS J002240.91+143110.4, which  is  a Luminous Red  Galaxy
(LRG),  and SDSS J002240.78+143113.9,   which   is  a very blue    and
elongated  object.    The   SDSS  targeted   SDSS  J002240.91+143110.4
(hereafter,  ``the LRG''; for LRG   selection, cf.\ Eisenstein et al.\
2001)  with a 3\arcsec   spectroscopic fiber.  The resulting  spectrum
(Fig.\   2) shows  absorption features   of   an early type galaxy  at
redshift of z=0.38 with Ca  H and K lines   at 5463 and 5510\AA.   The
very blue and elongated SDSS J002240.78+143113.9 was not targetted for
SDSS spectroscopy and hence has no SDSS spectrum.

Allam recognized this system as a probable gravitational lens and, due
to its time of discovery, dubbed it the ``8 o'clock arc.'' The arc is
a very blue high surface brightness object north of the LRG,
subtending an angle of $\approx$162\arcdeg about the galaxy.

The arc   consists of three components,   which are the  blue A1 (SDSS
J002240.78+143113.9) the reddish  blue A2  (SDSS J002240.96+143113.9),
and the blue A3 (SDSS J002241.14+143112.7).   The arc containing these
three components extends over 9.6\arcsec in length and has a length to
width ratio of 6:1.

\begin{figure}
\vspace{-.25cm}
\centering
\makebox[60mm]{\psfig{file=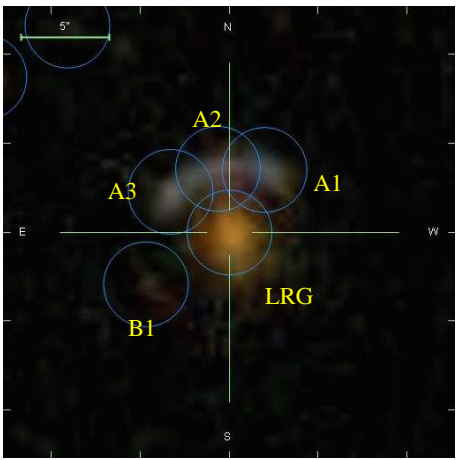,width=60mm,height=60mm,clip=,angle=0,silent=}}
\caption{SDSS $gri$ color image showing the location of three lensed 
images (A1,  A2, A3), the position of the LRG, and a faint back 
ground LRG (B1) at photometric redshift of 0.53. \label{figgri}}
\end{figure}   

\begin{figure}
\centering
\vspace{-.25cm}
\makebox[60mm]{\psfig{file=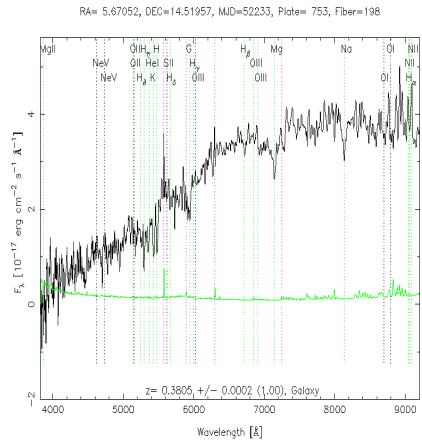,width=60mm,height=60mm,clip=,angle=0,silent=}}
\caption{The SDSS spectrum for the LRG SDSS J002240.91+143110.4. \label{figspec}}
\end{figure}  
\section{The Confirmation}\label{sec:conf}

In   order to   confirm  the   identification  of  the  system    as a
gravitational lens, we carried  out follow-up imaging and spectroscopy
on the Astrophysical Research  Consortium (ARC)  3.5m at Apache  Point
Observatory on the night of 2006 August 24 (UT).

\subsection{Imaging}

The   imaging was obtained  under  photometric  conditions  and with a
seeing of 1.0--1.2$\arcsec$ (FWHM) during the first half of the night.
The    instrument used  was   the  SPIcam  CCD imager,   which   has a
field-of-view  of 4.78$\arcmin\times4.78\arcmin$.   Three exposures of
300~sec each  were obtained  in  each of the    SDSS $gri$ filters;  a
15$\arcsec$ dithering pattern about the LRG was employed.

The  resulting images  were  processed  using the   IRAF  {\tt ccdred}
package.  The  images were then co-added with  the {\tt swarp} package
(Bertin  2000, v2.16;   see Fig.\    3),   and object  detection   and
measurement   were  made  with {\tt SExtractor}     (Bertin \& Arnouts
1996). We   used a weighted  coaddition,  accounting for  flux scaling
between the images, and    aperture  photometry with an   aperture  of
3$\arcsec$.  Photometric zeropoints were  derived by  matching objects
detected in the co-added images with objects  in the SDSS imaging data
and  comparing their {\tt   SExtractor}  {\tt MAG\_AUTO}  instrumental
magnitudes with their   SDSS model magnitudes.   The  $gri$ magnitudes
measured  from   these co-added images  are  listed  in  Table  1.  An
astrometric solution for the  coadded images was measured relative  to
the SDSS overlapping bright stars in the field of view.

\begin{figure}
\vspace{-.25cm}
\centering
\makebox[60mm]{\psfig{file=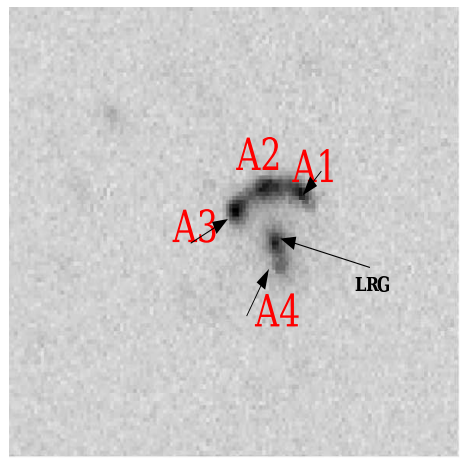,width=60mm,height=60mm,clip=,angle=0,silent=}}
\caption{
The coadded ARC 3.5m SPICAM $g$-band image clearly shows the three
components of the arc (A1, A2, A3) as well as the counter image (A4);
the center of the LRG is also marked. \label{figgri}}
\end{figure}   

\subsection{Spectroscopy}

Slit  spectroscopy was carried   out with the  DIS  III (Dual  Imaging
Spectrograph)  using  the standard Medium   Red/Low Blue grating setup
during  the second  half night  of night.  Six exposures were obtained
under moon-less conditions for a total exposure  time of 140 min.  The
seeing  was   1$\arcsec$--1.2$\arcsec$ (FWHM).      A slit  width   of
1.5\arcsec was employed, and the slit was oriented to cover as much of
the   three components of  the  8 o'clock arc   as  was possible.  The
standard Medium   Red/Low  Blue  grating  setup  covers   an effective
spectral range of 3600\AA\   to 9600\AA\  at  a linear  resolution  of
2.43\arcsec pix$^{-1}$   in  the  blue    part of the    spectrum  and
2.26\arcsec  pix$^{-1}$ in the   red; the spatial  scale is 0.4\arcsec
pix$^{-1}$.  HeNeAr  lamp    exposures   were taken  for    wavelength
calibration.  The   HST   spectrophotometric  standard   G191-B2B  was
observed for flux calibration.

The spectra were reduced using the IRAF {\tt ccdred} package and the
{\tt doslit} task.  The six individual spectroscopic exposures of the
8 o'clock arc were combined using the {\tt scombine} task, and the red
and blue spectra were spliced together using the {\tt spliceSpec} task
from Gordon Richard's {\tt distools} external IRAF package.

The redshift of the 8 o'clock arc was estimated to be $z=2.73$ based on 
measurements of 
Ly$\alpha$ $\lambda$ 1215.7,
SiII    $\lambda$1260.4,
OI+SiII $\lambda$(1302.2+1304.4), 
CII     $\lambda$1334.5, 
SiIV    $\lambda$1393.8, 
SiIV    $\lambda$1402.8,
SiII    $\lambda$1526.7,
CIV+CIV $\lambda$1549.5, and 
AlII    $\lambda$1670.8,
confirming that the arc is indeed a gravitational lens. Figure 4 shows
the spectrum of the 8 o'clock arc and, for comparison, the
SDSS-measured spectrum of the former ``brightest known LBG'' (cB58),
and an LBG composite spectra from Shapley et al.(2003).  

Both cB58  at $z=2.72$ and the 8  o'clock arc at  $z=2.73$ show damped
Ly$\alpha$, along with  the  typical stellar and  instellar absorption
lines found  in LBGs, quite similar  to those visible in the composite
spectrum.

\begin{figure}
\vspace{-.25cm}
\centering
\makebox[60mm]{\psfig{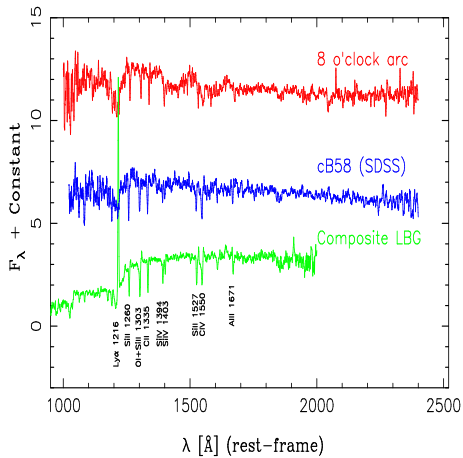}}
\caption{
The discovery spectrum taken with the DIS-III spectrograph on the ARC
3.5m telescope is plotted in red.  The SDSS spectrum of cB58, a
strongly lensed $L_{*}$ LBG at z = 2.72, is plotted in blue, and a
composite LBG spectra (Shapley et al.\ 2003) in green.  Several
important absorption lines are marked.
\label{figspec}}  
\end{figure}

\section{The Lens Model}\label{sec:model} 

\begin{figure}      
\centering
\vspace{-3.75cm}      
\makebox[80mm]{\psfig{file=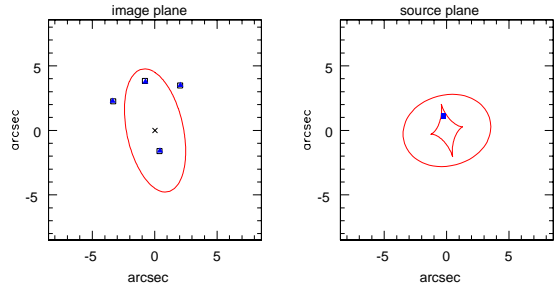,width=80mm,height=80mm,clip=,angle=0,silent=}}
\caption{
The results of fitting  an SIE model.   In the left plot  the measured
postions are shown as  blue triangles, the   fitted positions are  the
open boxes and the cross  represents the lensing  galaxy. In the right
plot the blue square represents the postion of the source. The critcal
curves and caustics are shown in red.  \label{figmodel}} 
\end{figure}

We began  with a simple mass  model to  reconstruct the lensing plane.
We used the pixel positions of the three lensed images and the counter
image  in  the coadded SPIcam  $g$  band  images  as measured  by {\tt
SExtractor}, since the  counter image (A4) is  not resolved  in either
the $r$ or $i$ band coadded  images.  The measured positions are shown
as blue   triangles in Figure   \ref{figmodel}.    We then assumed   a
singular isothermal ellipse   (SIE), and used the {\tt  gravlens}/{\tt
lensmodel}  software of Keeton  (2001)  to perform fits; the resulting
fitted  position are plotted in  Figure \ref{figmodel}  as open boxes.
The best fit SIE model yields an Einstein  radius of $\theta_{\rm Ein}
= 2.91\arcsec\pm0.14\arcsec$,   or $R_{\rm  Ein} = 10.6\pm0.5  h^{-1}$
kpc.   The best fit  $\chi^{2}$  is 2.1 for   NDF=7, where we  assumed
positional errors of $\pm$0.1\arcsec.
  
Since both the redshift of the LRG and the  LBG are known we were able
to determine the angular diameter distance to the source ($D_{s}$), to
the lens ($D_{l}$), and between the source and  lens ($D_{sl}$), to be
1141, 752 and 863 $h^{-1}$ Mpc, respectively.  The total magnification
was found to be a factor  of $12.3^{+15}_{-3.6}$ ($\ga  4$ for each of
the  three   arc images).   The   velocity  dispersion  of  the   mass
distribution  doing   the lensing   was  predicted    to be  $391  \pm
19$~km~s$^{-1}$,  which   is large   but   not unprecedented   for  an
elliptical galaxy.  (E.g., Crampton et  al.  2002 modelled a  velocity
disperson of  387$\pm$5~km~s$^{-1}$ for the strongly lensed elliptical
galaxy CFRS~03.1077, and Bernardi  et al.\ 2006 find $\sim$50 galaxies
with  $\sigma >  350$~km~s$^{-1}$    from  a SDSS  sample   of  39,320
elliptical galaxies.)

The  fitted ellipticity and position  angle  are $0.53\pm0.06$ and $12
\arcdeg \pm 2\arcdeg$,  respectively.  These fitted values agreed well
within 1$\sigma$ with the observed values for these parameters for the
LRG in the SDSS DR4 database (0.46 and 12$\arcdeg$, respectively).

From  this  simple SIE  model we  can  determine the  mass interior to
$R_{\rm  Ein}$  using   $M_{\rm Ein}=  (c^{2}/4G)(D_{l}  D_{s}/D_{sl})
\times   \theta_{\rm     Ein}^{2}$.     We     find     that   $M_{\rm
Ein}=1.04\times10^{12} h^{-1} M_{\sun}$.  As we know the lens distance
we can determine the mass-to-light  ratio and we  find a value of $16h
M_{\sun}/L_{\sun}$ ($i$-band).


\section{Discussion}\label{sec:discussion}

The  spectrum  of the 8  o'clock  arc (Fig. 4)   shows that the lensed
source galaxy is an  LBG, albeit an uncommonly  bright one: lensed, it
is  4.6  mag brighter ($g$-band)  than   an $L_*$ LBG  (Adelberger  \&
Steidel  2000).  Even after  accounting for a lensing magnification of
$\approx 12.3$ (\S~\ref{sec:model}), the   8  o'clock arc is  1.9  mag
($\approx$ a factor of 6) more luminous than $L_*$ for LBGs.

For comparison,  cB58 is  a  typical  $L_*$ LBG   lensed by the  large
$z=0.37$  foreground  cluster    MS  1512+36   (Yee et   al.\   1996).
Furthermore,  cB58 is magnified  by a factor of  $\sim$  30 and has an
apparent  brightness only about  one-third that  of  the 8 o'clock arc
(Seitz  et al.\ 1998).  We also  note that  the relative simplicity of
the environment surrounding the 8 o'clock  arc's lensing LRG permits a
quite robust determination  of the lensing amplification, whereas  the
lensing amplification   for cB58 is  rather  sensitive  to the assumed
cluster mass distribution model.

We can also estimate the star formation rate  of the 8 o'clock arc LBG
using a scaling relation given in  Pettini et al (2000).  The relation
is given for cB58 but for the accuracy necessary here we can take cB58
and the 8 o'clock arc to be at the same redshift:
${\rm SFR} = 3 \times 37\times (\frac{30}{f_{lens}})\times(\frac{f_{dust}}{7})
    \times (\frac{2.5}{f_{\rm IMF}}) M_\sun {\rm yr}^{-1}$
where  the additional factor of 3   over the Pettini  et al.\ relation
(their eq.\ [6]) takes into account the fact the 8 o'clock arc's total
apparent brightness is roughly  three times that  of cB58.  For the  8
o'clock arc, $f_{lens} = 12.3$ and we take  the other parameters to be
the same, leading to   SFR $= 270  M_\sun  {\rm yr}^{-1}$ ($H_0$  = 70
km~s$^{-1}$~Mpc$^{-1}$ and $q_0$=0.1), or   SFR $= 160  h^{-1}  M_\sun
{\rm yr}^{-1}$ ($\Omega_{\rm  m}= 0.3$, $\Omega_{\lambda}=0.7$).  This
should be taken as an estimate only, as the Pettini et al. relation is
based upon UV  continuum luminosity and gives  the highest of  all the
star formation rate estimates for  cB58.  Followup measurements of the
8 o'clock  will  provide more detailed   rate estimates.   Taking  the
estimate at face value, the 8 o'clock arc is in the top $20\%$ of star
formation rates given for LBGs in Shapley et al.\ (2001).

Understanding the  environment  of  the  LRG  and  the  lensed LBG  is
critical    for interpreting the     results of the gravitational lens
modeling.   Therefore, we have  run maxBcg cluster finder (Koestler et
al.\ 2006) on  the area.  The  LRG sits near  a maxBcg cluster, with a
center       at     ($\alpha$,$\delta$)$_{\rm        J2000}$         =
(5.6749\arcdeg,14.5003\arcdeg)  and a  photometric redshift of $z_{\rm
photo} = 0.384$  and richness $N_{gals}  = 21$.  This  is a reasonably
massive  cluster, and the  lensing  LRG sits  just  260  kpc  from the
brightest  cluster galaxy.  This may give  an additional sheet surface
mass contribution to the lensing, but it is likely that this is a very
small effect (Kochanek, private communication).

\section{Conclusions}\label{sec:conclusions}

We have  reported on  the discovery of   a  strongly lensed LBG  at  a
redshift of $z=2.73$, the  arc of which we  have named the ``8 o'clock
arc.''   At   an    apparent     magnitude    of ($g$,$r$,$i$)     $=$
(19.95,19.22,18.98),  it displaces cB58 as  the brightest known LBG by
over a magnitude.

The  arc consists of three   lensed images of  the  LBG and subtends a
length of 162\arcdeg (9.6\arcsec)  around the  the lensing galaxy,  an
early type galaxy at $z=0.38$.   The length-to-width ratio of the  arc
is 6:1.  A fourth (counter) image is  also visible in the co-added ARC
3.5m $g$-band image.

A simple SIE  lens model for the system  yields an  Einstein radius of
$\theta_{\rm   Ein}=2.91\arcsec\pm0.14\arcsec$     ($R_{\rm      Ein}=
10.6\pm0.5$ $h^{-1}$~kpc), a total  lensing  mass within the  Einstein
radius  of $1.04\times10^{12}  h^{-1} M_{\sun}$,   and a magnification
factor for the  LBG of $12.3^{+15}_{-3.6}$.   Based  upon this model's
value for the  magnification factor, it is clear  that the LBG is  not
only apparently bright but also quite intrinsically luminous (about $6
\times L_{*}$).  Furthermore, a  simple scaling relation  from Pettini
et al.\ (2000) indicates that  the LBG may  be experiencing an episode
of vigorous  star formation, perhaps as high  as 160 $h^{-1} M_{\sun}$
yr$^{-1}$.  The remarkably low apparent magnitude of this object makes
it an     ideal  object for  further  followup    with   more detailed
observations.


\begin{deluxetable}{@{\,}c@{\,}c@{\,}c@{\,}c@{\,}c@{\,}c@{\,}c@{\,}c@{\,}}
\tablecaption{APO 3.5m SDSS Photometry.\label{tabin2}}
\tablewidth{0pt}
\tablenum{1}
\tabletypesize{\tiny}
\tablehead{
\colhead{ID}& 
\colhead{ra (2000.0)} & 
\colhead{dec (2000.0)} & 
\colhead{g\tablenotemark{\ddag}} &
\colhead{r\tablenotemark{\ddag}} &  
\colhead{i\tablenotemark{\ddag}} 
}
\startdata
LRG      & 00:22:40.91 & 14:31:10.0 & 20.14& 18.62& 18.16\\
A1       & 00:22:40.79 & 14:31:13.8 & 21.18& 20.21& 20.13\\
A2       & 00:22:40.97 & 14:31:14.0 & 20.99& 20.40& 20.11\\
A3       & 00:22:41.15 & 14:31:12.6 & 21.27& 20.68& 20.21\\
B1       & 00:22:41.44 & 14:31:06.8 & 23.67& 23.02& 22.40\\
Arc total mag   &             &            & 19.95& 19.22& 18.96 \\
Arc $\mu$mag arcsec$^{-2}$  &             &            & 23.26& 22.54& 22.28 \\
A4       & 00:22:40.89 & 14:31:08.9 & $\sim$22 & \nodata & \nodata \\ 
\enddata
\tablenotetext{\ddag}{The estimated magnitude error is $\pm$0.1 mag. The magnitudes 
listed have not been corrected for instellar extinction, which is 0.22
mag, 0.16 mag, and 0.12 mag, for $g$, $r$, and $i$, respectively
(Schlegel et al.\ 1998). }
\end{deluxetable}

\acknowledgments    

We thank Chris Kochanek for useful discusions on the lensing model.

S.S.A. acknowledges support from NSF NVO Grant No. AST-0122449.

These results are based on observations obtained with the Apache Point
Observatory 3.5-meter telescope, which  is owned  and operated by  the
Astrophysical Research Consortium.

Funding  for the SDSS  and  SDSS-II has been  provided  by  the Alfred
P.  Sloan Foundation,   the  Participating Institutions, the  National
Science   Foundation,  the  U.S. Department   of  Energy, the National
Aeronautics and Space Administration, the Japanese Monbukagakusho, the
Max Planck  Society, and  the  Higher  Education Funding  Council  for
England. The SDSS Web Site is http://www.sdss.org/.


\end{document}